\documentstyle[12pt]{article}

\newcommand{\be}{\begin{equation}} 
\newcommand{\ee}{\end{equation}} 
\newcommand{\bea}{\begin{eqnarray}} 
\newcommand{\eea}{\end{eqnarray}} 
 
\newcommand{\p}[1]{(\ref{#1})} 
  
\begin{document}
\begin{flushright}
hep-th/9601173\\
January 1996
\end{flushright}
\centerline{\large\bf $SU(2)\times SU(2)$ harmonic superspace 
and }
\vspace{0.2cm}
\centerline{\large\bf (4,4) sigma models with torsion}
\vskip0.4cm
\centerline{{\large Evgeny A. Ivanov}}
\vskip.2cm
\centerline{\it Bogoliubov Laboratory of Theoretical Physics, JINR,}
\centerline{\it 141 980, Dubna, Moscow Region, Russia}
\vskip.3cm
\begin{center}
Talk at the 29th International Symposium on the Theory 
of Elementary Particles, \\
August 1995, Buckow, Germany 
\end{center}

\vskip.3cm

\begin{abstract}\noindent{\small
We review a manifestly supersymmetric off-shell 
formulation of a wide class of torsionful $(4,4)$ $2D$ sigma models 
and their massive deformations in the harmonic superspace 
with a double set of $SU(2)$ harmonic variables. Sigma models 
with both commuting and non-commuting left and right complex structures 
are treated.}  

\end{abstract}
\vskip.3cm

\noindent{\bf 1. Introduction.}
Manifestly supersymmetric off-shell superfield formulations 
of $2D$ sigma models with extended worldsheet SUSY are most 
appropriate for revealing remarkable target geometries of these 
theories. For torsionless $(2,2)$ and $(4,4)$ 
sigma models the relevant superfield Lagrangians coincide 
with (or are directly related to) the 
fundamental objects underlying the bosonic target geometry: 
K\"ahler potential in the $(2,2)$ case \cite{Zum},  
hyper-K\"ahler or quaternionic-K\"ahler potentials in the flat or 
curved $(4,4)$ cases [2 - 5]. One of the basic advantages of such a 
description is the possibility to explicitly compute the corresponding 
bosonic metrics (K\"ahler, hyper-K\"ahler, quaternionic ...) starting 
from an unconstrained superfield action \cite{{GIOShk},{DVal}}. This is 
to be compared with the approach when only $(1,1)$ supersymmetry is kept 
manifest, while the restrictions imposed by higher extended (on-shell)
supersymmetries amount to the existence of an appropriate 
number of covariantly constant complex structures on the bosonic 
manifold. Such an approach yields no explicit recipes for 
computing the bosonic metrics. To have superfield off-shell formulations 
with all supersymmetries manifest is also highly desirable while quantizing 
these theories and proving their ultraviolet finiteness.   

An important class of $2D$ supersymmetric sigma models 
is given by $(2,2)$ and $(4,4)$ models with torsionful bosonic target 
manifolds and two 
independent left and right sets of complex structures (see, 
e.g. \cite{{GHR},{HoPa}}). 
These models and, 
in particular, their group manifold WZNW representatives \cite{belg} 
can provide non-trivial 
backgrounds for 
$4D$ superstrings (see, e.g., \cite{Luest}) and be relevant to 
$2D$ black holes in the stringy context \cite{{RSS},{CHS}}.  
A manifestly supersymmetric formulation of $(2,2)$ models with 
commuting left and right complex structures in terms of chiral and 
twisted chiral $(2,2)$ 
superfields and an exhaustive discussion of their geometry have been 
given in \cite{GHR}. For $(4,4)$ models with commuting structures there 
exist manifestly supersymmetric off-shell formulations 
in the projective, ordinary and $SU(2)\times SU(2)$ harmonic $(4,4)$ 
superspaces [11,13,14]. The appropriate superfields represent, in one 
or another way, the $(4,4)$ $2D$ twisted 
multiplet \cite{{IK},{GHR}}. 

Much less is known about $(2,2)$ and $(4,4)$ sigma models with 
non-commuting 
complex structures, including most of group manifold ones \cite{belg}. 
In particular, it is unclear how 
to describe them off shell in general. As was argued in Refs. 
\cite{{GHR},{BRL},{RSS}}, twisted $(2,2)$ and $(4,4)$ multiplets are 
not suitable for this purpose. It has been then suggested to make use 
of some other off-shell representations of $(2,2)$ \cite{{BRL},{DS}} 
and $(4,4)$ \cite{{BRL},{RIL}} worldsheet SUSY. However, it is an open 
question whether the relevant actions correspond to generic sigma models of 
this type. 

In this talk we review another approach to the off-shell 
description of general $(4,4)$ sigma models with torsion and their 
massive deformations, largely exploiting an analogy with general torsionless 
hyper-K\"ahler $(4,4)$ sigma models in $SU(2)$ harmonic superspace [2 - 4]. 
The basic tool is the new type of harmonic superspace containing two 
independent sets of harmonic variables, the $SU(2)\times SU(2)$ harmonic 
superspace. We demonstrate that it allows to construct 
off-shell formulations for a wide class of torsionful $(4,4)$ sigma models 
with commuting as well as non-commuting left and right quaternionic 
structures \cite{{IS},{Iv},{Iv2}}. 

\vspace{0.2cm}
\noindent{\bf 2. (4,4) twisted multiplet in SU(2)xSU(2) 
harmonic superspace.}
The $SU(2)\times SU(2)$ harmonic superspace is an extension of the 
standard real $(4,4)$ $2D$ superspace by two independent sets 
of harmonic variables  $u^{\pm 1\;i}$ and $v^{\pm 1\;a}$ 
($u^{1\;i}u^{-1}_{i} =
v^{1\;a}v^{-1}_{a} = 1$) associated with 
the automorphism groups $SU(2)_L$ and $SU(2)_R$ of the left and 
right sectors of $(4,4)$ supersymmetry \cite{IS}. The corresponding 
analytic subspace is spanned by the following set of coordinates 
\be  \label{anal2harm}
(\zeta, u,v) = 
(\;x^{++}, x^{--}, \theta^{1,0\;\underline{i}}, 
\theta^{0,1\;\underline{a}}, u^{\pm1\;i}, v^{\pm1\;a}\;)\;,  
\ee
where we omitted the light-cone indices of odd coordinates (the first and 
second $\theta$s in \p{anal2harm} carry, respectively, the indices $+$ and 
$-$). The superscript ``$n,m$'' stands for two independent harmonic $U(1)$ 
charges, left ($n$) and right ($m$) ones. The additional doublet indices 
of odd coordinates, $\underline{i}$ and $\underline{a}$, refer to 
two extra automorphism groups $SU(2)_L^{'}$ and $SU(2)_R^{'}$ which 
together with $SU(2)_L$ and $SU(2)_R$ form the full $(4,4)$ $2D$ 
supersymmetry automorphism group $SO(4)_L \times SO(4)_R$. 

It was argued in \cite{IS} that this type of harmonic superspace is most 
appropriate for constructing off-shell formulations of general $(4,4)$ 
sigma models with torsion. This hope mainly relied upon the fact 
that the twisted $(4,4)$ multiplet has a natural description as a 
real analytic $SU(2)\times SU(2)$ harmonic superfield $q^{1,1}(\zeta,u,v)$ 
subjected to the harmonic constraints
\be  \label{qconstr}
D^{2,0} q^{1,1} = D^{0,2} q^{1,1} = 0\;.
\ee
where 
\bea
D^{2,0} &=& \partial^{2,0} + i\theta^{1,0\;\underline{i}}
\theta^{1,0}_{\underline{i}}\partial_{++}\;, 
\;\;D^{0,2} \;=\;\partial^{0,2} + i\theta^{0,1\;\underline{a}}
\theta^{0,1}_{\underline{a}}\partial_{--}  
\label{harm2der} \\
(\partial^{2,0} &=& u^{1 \;i}\frac{\partial} {\partial u^{-1 \;i}}\;,\;\; 
\partial^{0,2} \;=\; v^{1 \;a}\frac{\partial} {\partial v^{-1 \;a}}) 
\nonumber 
\eea
are the left and right mutually commuting analyticity-preserving 
harmonic derivatives. These constraints leave in $q^{1,1}$ $8+8$ 
independent components that is just the irreducible off-shell 
component content 
of $(4,4)$ twisted multiplet. The 
most general off-shell action of $n$ such multiplets $q^{1,1\;M} (M=1,...n)$ 
is given by the following integral over the analytic superspace 
(\ref{anal2harm})  
\be
S_{q} = \int \mu^{-2,-2}\; h^{2,2}(q^{1,1\;M},u^{\pm1},v^{\pm1})\;,  
\label{qact}
\ee
$\mu^{-2,-2}$ being the relevant integration measure. 
The analytic superfield lagrangian $h^{2,2}$ is an arbitrary function 
of its arguments (consistent with the external $U(1)$ harmonic 
charges $2,2$).  

Let us write down the physical bosons part of the action \p{qact}
\be
S_{phb} = {1\over 2} \int d^2 z\; \{ G_{Mia\;Njb} (q) \;\partial_{++}
q^{ia\;M} 
\partial_{--}q^{jb\;N} + B_{Mia\;Njb}(q)\; \partial_{++}
q^{ia\;M} 
\partial_{--}q^{jb\;N} \}\;.
\label{compbosph}
\ee
Here 
\bea
G_{Mia\;Njb} (q) &=& G_{M\;N} (q)\; \epsilon_{ij} \epsilon_{ab}\;, 
\label{metr} \\ 
B_{Mia\;Njb}(q) &=& \int du dv\; g_{M\;N}(q^{1,1}_0, u, 
v)[\epsilon_{ij} v^{1}_{(a} v^{-1}_{b)} - 
\epsilon_{ab} u^{1}_{(i} u^{-1}_{j)}]\;, \label{tors} \\
G_{M\;N} (q) &=& \int du dv\; g_{M\;N}(q^{1,1}_0, u, v)\;, \;\;\;
g_{M\;N}(q^{1,1}_0, u, v) = 
\frac{\partial^2 h^{2,2}}{\partial q^{1,1\;M} 
\partial q^{1,1\;N}}|_{\theta = 0}\;,
\label{auxfun} 
\eea
where $q_0^{1,1} \equiv q^{1,1} |_{\theta = 0}$. The 
symmetric and skew-symmetric objects $G_{Mia\;Njb}$ and 
$B_{Mia\;Njb}$ can be identified with the metric and torsion 
potential on the target space. 

It is advantageous to rewrite the second 
term in (\ref{compbosph}) through the torsion field strength:  
\be
H_{Mia\;Njb\;Tkd} = \partial_{Mia} B_{Njb\;Tkd} +
\partial_{Tkd} B_{Mia\;Njb}+
\partial_{Njb} B_{Tkd\;Mia} \;,
\label{hgen}
\ee
where $\partial_{Mid} \equiv \partial / \partial q^{id\;M}$. 
Letting $q^{ia\;M}$ depend on an extra parameter 
$t$, with $q^{ia\;M}(t)|_{t=1} \equiv q^{ia\;M}$, 
$q^{ia\;M}(t)|_{t=0} = \epsilon^{ia}$, one can locally 
rewrite the torsion term as
\be
B_{Mia\;Njb}\; \partial_{++} q^{ia\;M} \partial_{--} q^{jb\;N} = 
\int^1_0 dt\; H_{Mia\;Njb\;Tkd}\; \partial_t q^{ia\;M} 
\partial_{++} q^{jb\;N} \partial_{--} q^{kd\;T}\;.
\label{torsHg}
\ee
For $B_{Mia\;Njb}$ given by eq. (\ref{tors}), 
$H_{Mia\;Njb\;Tkd}$ is reduced to   
\be 
H_{Mia\;Njb\;Tkd}(q) = \partial_{(Mid} \; G_{N\;T)}(q) \;
\epsilon_{ab}
\epsilon_{jk} + \partial_{(Mka}\; G_{N\;T)}(q)\; 
\epsilon_{db} \epsilon_{ij}\;.
\label{exprHg}
\ee
Note that all the fermionic terms in the action 
(\ref{qact}) are also 
expressed through the same function $G_{M\;N}(q)$ and its 
derivatives. 

In the case of four-dimensional targets 
(the case of one $q^{1,1}$) the metric, as is seen from eqs. \p{metr} - 
\p{auxfun}, is reduced to a conformal factor
\be
G_{ia\;jb} (q) =  \epsilon_{ij} \epsilon_{ab} G(q) \equiv 
\epsilon_{ij} \epsilon_{ab} \int du dv\;g(q_0^{1,1}, u,v)\;, \;\;\; 
g = \frac{\partial^2 }{\partial q_0^{1,1} 
\partial q_0^{1,1}} h^{2,2}(q_0^{1,1}, u,v) \label{conffl}\;
\ee 
which satisfies the Laplace equation 
\be
\Box G(q) \equiv \partial^{ia} \partial_{ia}
G(q) = 2 \int du dv \left( \frac{\partial^2}{\partial q_0^{-1,-1} 
\partial q_0^{1,1}} - \frac{\partial^2}{\partial q_0^{-1,1} 
\partial q_0^{1,-1}} \right) g(q_0^{1,1}, u,v) = 0.  \label{laplace}
\ee
This agrees with the general conditions on the bosonic target in the 
torsionful $(4,4)$ sigma models with four-dimensional targets 
\cite{{GHR},{CHS}}.

As a non-trivial example of the $q^{1,1}$ action with 
four-dimensional bosonic manifold we give 
the action of $(4,4)$ extension of the $SU(2)\times U(1)$ 
WZNW sigma model  
\be \label{wzwaction} 
S_{wzw} = \frac{1}{4\kappa ^2} \int \mu^{-2,-2} \;\hat{q}^{1,1} 
\hat{q}^{(1,1)} 
\left(\frac{\mbox{ln}(1+X)}{X^2} - \frac{1}{(1+X)X} \right)\;. 
\label{confact} 
\ee 
Here  
\be 
\hat{q}^{1,1} = q^{1,1} - c^{1,1}\;,\;X = c^{-1,-1}\hat{q}^{1,1}\;,\; 
c^{\pm 1,\pm 1} = c^{ia}u^{\pm1}_iv^{\pm1}_a \;,\; 
c^{ia}c_{ia} = 2\;. 
\ee
Despite the presence of an extra quartet constant $c^{ia}$ in the 
analytic superfield lagrangian, the action (\ref{wzwaction}) 
actually does not depend 
on $c^{ia}$ \cite{IS} as it is invariant under arbitrary rigid
rescalings and $SU(2)\times SU(2)$ rotations of this constant. 
Its physical bosons part is given by the general expression 
\p{compbosph} and eventually turns out to be expressed through 
the single function $G(q)$ which in the case under consideration 
reads, up to the overall coupling constant,
\be
G(q) = \int du dv \frac{1 - X}{(1+X)^3} = 2 (q^{ia} q_{ia})^{-1} = 
e^{-2u} \;,
\ee 
where we have introduced the singlet scalar field $u(x)$ through the polar 
decomposition 
\be
q^{ia} = e^{u(z)} \tilde{q}^{ia}(z)\;,\;\;
\tilde{q}^{ia} \tilde{q}^{j}_{\;a} = \epsilon^{ji}\;, 
\tilde{q}^{ia} \tilde{q}_{i}^{\;b} = \epsilon^{ba}\;. 
\label{newparam}
\ee
In this parametrization, the resulting action of physical bosons 
is that of the $SU(2)\times U(1)$ WZNW sigma model 
\bea
S_{wzw}^{bos} &=& {1\over 4\kappa^2} \int d^2 z \;\{ 
\partial_{++}u \partial_{--}u + {1\over 2} 
\partial_{++}\tilde{q}^{ia} 
\partial_{--}\tilde{q}_{ia} \nonumber \\
&& + 
{1\over 2} \int^1_0 dt \; \partial_t\tilde{q}_{ia}\;
\tilde{q}_{jb}\; (
\partial_{++}\tilde{q}^{ib}\partial_{--}\tilde{q}^{ja} - 
\partial_{++}\tilde{q}^{ja}\partial_{--}\tilde{q}^{ib}) \}\;. 
\label{okonb}
\eea
The fermionic and auxiliary fields parts of the action \p{wzwaction}
also have the appropriate form \cite{IS}. 

The last topic we wish to discuss in connection with the action \p{qact} 
concerns massive deformations of the latter. Surprisingly, and this 
is a crucial difference of the considered $(4,4)$ case from, say, $(2,2)$ 
sigma models with torsion, the only massive term of $q^{1,1\;M}$ 
consistent with analyticity and off-shell $(4,4)$ supersymmetry (not 
modified by central charges) is the following one \cite{{GI},{IS}}
\be 
S_m = m \int \mu^{-2,-2}\; 
\theta^{1,0\;\underline{i}} 
\theta^{0,1\;\underline{b}}\; 
C_{\underline{i}\;\underline{b}}^M\; 
q^{1,1\;M}
\;; \;\; [m] = cm^{-1}\;,
\label{mtermg}
\ee
where $C_{\underline{i}\;\underline{b}}^M $ are arbitrary 
constants (subject to the appropriate reality conditions for the 
modified action to remain real). 
It immediately follows that, despite the presence of 
explicit $\theta$'s, (\ref{mtermg}) 
is invariant under rigid $(4,4)$ 
supersymmetry: one represents the supertranslation of, say, 
$\theta^{1,0\;\underline{i}}$ as 
$$
\delta_{SUSY}\; \theta^{1,0\;\underline{i}} = 
\epsilon^{k\underline{i}} u^{1}_k 
= D^{2,0} \epsilon^{k\underline{i}} u^{-1}_k\;, 
$$
integrates by parts with respect to $D^{2,0}$ and makes use 
of the defining constraints (\ref{qconstr}). 

After adding the term \p{mtermg} to the action \p{qact}, passing to 
components and eliminating auxiliary fields, the effective addition to the 
$(4,4)$ sigma model component action is given by 
\be
S_{q}^{pot} = {{m^2}\over 2} \int d^2 z \;G^{M\;N}(q)\;
(C_{\underline{i}\;\underline{a}}^M C^{\;\underline{i}\;
\underline{a}\;N}) \;,
\ee 
where $G^{M\;N}(q)$ is the inverse of the metric $G_{M\;N}(q)$ defined  
in eq. \p{auxfun}. Thus we see that the potential term in the case in 
question is uniquely determined by the 
form of the bosonic target metric. In particular, in the case of $(4,4)$ 
$SU(2)\times U(1)$ WZNW model one gets the Liouville potential term for 
the field $u(x)$, so the massive deformation of this model is nothing but 
the $(4,4)$ super Liouville theory \cite{IK}. It would be interesting to 
inquire whether $(4,4)$ extensions of other integrable $2D$ theories 
(e.g., sine-Gordon theory) can be obtained as massive deformations of 
some appropriate torsionful $(4,4)$ sigma models.

Note that only for non-trivial curved bosonic 
targets (with non-trivial dependence on $q^{1,1}$ in $G^{M\;N}$) the above 
term actually produces a mass for $q^{1,1}$. One cannot 
gain a mass for $q^{1,1}$ in this way, starting with the free lagrangian 
$h^{2,2} \sim q^{1,1\;M} q^{1,1\;M}$ (corresponding to the $U(1)^n$ 
bosonic target). This becomes possible only after including 
into the game another type of $(4,4)$ twisted multiplet which has no 
simple description in the considered type of $(4,4)$ harmonic 
superspace \cite{{GI},{GK}}.  

\vspace{0.2cm}
\noindent{\bf 3. More general (4,4) sigma models with torsion.}
The above $SU(2)\times SU(2)$ harmonic superspace description of 
$(4,4)$ twisted multiplet suggests a new off-shell formulation  
of the latter via unconstrained analytic superfields. After implementing  
the constraints \p{qconstr} in the action with superfield lagrange 
multipliers and adding this term to \p{qact} we arrive at 
the following new action \cite{IS}
\be
S_{q,\omega} = \int \mu^{-2,-2} \{ 
q^{1,1\;M}(\;D^{2,0} \omega^{-1,1\;M}   + 
D^{0,2}\omega^{1,-1\;M} \;) + h^{2,2} (q^{1,1}, u, v) \}\;.
\label{dualq}
\ee
In (\ref{dualq}) all the involved superfields are unconstrained analytic, 
so from the beginning the action (\ref{dualq}) contains an infinite number 
of auxiliary fields coming from the double harmonic expansions with 
respect to the harmonics $u^{\pm1\;i}, v^{\pm1\;a}$. Varying 
with respect to the Lagrange multipliers $\omega^{1,-1\;M}, 
\omega^{-1,1\;M}$ takes one back to the action \p{qact} and 
constraints \p{qconstr}. On the other hand, varying with respect to 
$q^{1,1\;M}$ yields an algebraic equation for the latter which allows 
one to get 
a new dual off-shell representation of the twisted multiplet action through 
unconstrained analytic superfields $\omega^{-1,1\;M}$, $\omega^{1,-1\;M}$. 

The crucial feature of the action (\ref{dualq}) (and its $\omega$ 
representation) is the abelian gauge invariance 
\be  
\delta \;\omega^{1,-1\;M} = D^{2,0} \sigma^{-1,-1\;M} 
\;, \; \delta \;\omega^{-1,1\;M}  
= - D^{0,2} \sigma^{-1,-1\;M}\;,
\label{gauge}
\ee
where $\sigma^{-1,-1\;M}$ are unconstrained analytic 
superfield parameters. This gauge freedom ensures the 
on-shell equivalence of the $q,\omega$ or $\omega$ formulations of 
the twisted multiplet action to its original $q$ formulation 
\p{qact}: it 
neutralizes superfluous physical dimension component fields in the 
superfields $\omega^{1,-1\;M}$ and $\omega^{-1,1\;M}$ and 
thus equalizes the number of propagating fields in both formulations. It 
holds already at the free level, with 
$h^{2,2}$ quadratic in $q^{1,1\;M}$, so it is natural to expect that 
any reasonable generalization of the action (\ref{dualq}) respects 
this symmetry or a generalization of it. We will see that this is 
indeed so. 
  
The dual twisted multiplet action (\ref{dualq}) is 
a good starting point for constructing more general actions which 
embrace sigma models with non-commuting left and right 
complex structures. 

As a natural generalization of \p{dualq} we insert an arbitrary 
dependence on $\omega^{-1,1\;M}$, $\omega^{1,-1\;M}$ 
into $h^{2,2}$. In other words, as an ansatz 
for the general action we take the following one 
\be
S_{gen} = \int \mu^{-2,-2} \{ 
q^{1,1\;M}(\;D^{2,0} \omega^{-1,1\;M}   + 
D^{0,2}\omega^{1,-1\;M} \;) + H^{2,2} (q^{1,1}, \omega^{1,-1}, 
\omega^{-1,1}, 
u, v) \}\;,
\label{genact}
\ee
where for the moment the $\omega$ dependence in $H^{2,2}$ is not fixed. 
In refs. \cite{{Iv},{Iv2}} we have shown that one can arrive at 
this ansatz proceeding from the most general form of $q,\omega$ action. 

It turns out that the $\omega$ dependence of the potential 
$H^{2,2}$ in (\ref{genact}) is actually completely specified by the 
integrability conditions following from the commutativity condition  
\be \label{comm}
[\;D^{2,0}, D^{0,2}\;] = 0\;.
\ee

To show this, we first write the equations of motion corresponding to 
(\ref{genact})
\bea 
D^{2,0}\omega^{-1,1\;M} + D^{0,2}\omega^{1,-1\;M} &=& - 
\frac{\partial H^{2,2} (q,\omega,u,v)}{\partial q^{1,1\;M}}\;, 
\label{eqom} \\
D^{2,0}q^{1,1\;M} \;=\; 
\frac{\partial H^{2,2} (q,\omega,u,v)}{\partial \omega^{-1,1\;M}}\;, 
\;\;
D^{0,2}q^{1,1\;M} &=& 
\frac{\partial H^{2,2} (q,\omega,u,v)}{\partial \omega^{1,-1\;M}}\;.
\label{eqqu}
\eea 
Applying the integrability condition (\ref{comm}) to the pair of 
equations (\ref{eqqu}) and making a natural assumption that it 
is satisfied as a consequence of the equations of motion (i.e. does not 
give rise to any new dynamical restrictions), after some algebra 
we arrive at the set of self-consistency conditions \cite{{Iv},{Iv2}}. 
One of their consequences is the following restriction on the $\omega$ 
dependence of $H^{2,2}$ 
\bea
H^{2,2} &=& h^{2,2}(q,u,v) + \omega^{1,-1\;N} h^{1,3\;N}(q,u,v) 
+ \omega^{-1,1\;N} h^{3,1\;N}(q,u,v) \nonumber \\
&& + \;\omega^{-1,1\;N}\omega^{1,-1\;M}
h^{2,2\;[N,M]}(q,u,v)\;. \label{Hgen}
\eea
Plugging this expression back into the self-consistency relations, 
one finally deduces four independent constraints on the potentials 
$h^{2,2}$, $h^{1,3\;N}$, $h^{3,1\;N}$ and $h^{2,2\;[N,M]}$
\bea 
&& \nabla^{2,0} h^{1,3\;N} - \nabla^{0,2} h^{3,1\;N} + h^{2,2\;[N,M]} 
\;\frac{\partial h^{2,2}}{\partial q^{1,1\;M}} \;=\; 0 \label{1} \\
&& \nabla^{2,0} h^{2,2\;[N,M]} - 
\frac{\partial h^{3,1\;N}}{\partial q^{1,1\;T}} \;h^{2,2\;[T,M]} + 
\frac{\partial h^{3,1\;M}}{\partial q^{1,1\;T}}\; h^{2,2\;[T,N]} \;=\; 0 
\label{2} \\
&& \nabla^{0,2} h^{2,2\;[N,M]} - 
\frac{\partial h^{1,3\;N}}{\partial q^{1,1\;T}}\; h^{2,2\;[T,M]} + 
\frac{\partial h^{1,3\;M}}{\partial q^{1,1\;T}}\; h^{2,2\;[T,N]} \;=\; 0 
\label{3} \\
&& h^{2,2\;[N,T]}\;\frac{\partial h^{2,2\;[M,L]}}{\partial q^{1,1\;T}} + 
h^{2,2\;[L,T]}\;\frac{\partial h^{2,2\;[N,M]}}{\partial q^{1,1\;T}} +
h^{2,2\;[M,T]}\;\frac{\partial h^{2,2\;[L,N]}}{\partial q^{1,1\;T}} 
\;=\; 0\;, 
\label{4} 
\eea
\be
\nabla^{2,0} = \partial^{2,0} + h^{3,1\;N}\frac{\partial}{\partial 
q^{1,1\;N}}
\;,\;\; 
\nabla^{0,2} = \partial^{0,2} + h^{1,3\;N}\frac{\partial}{\partial 
q^{1,1\;N}}
\;.
\ee
Here $\partial^{2,0}, \partial^{0,2}$ act only on the ``target'' 
harmonics, i.e. those appearing 
explicitly in the potentials.

Thus the true analog of the generic 
hyper-K\"ahler $(4,4)$ sigma model action [2 - 4] in the torsionful case 
is the action 
\bea
S_{q,\omega} &=& 
\int \mu^{-2,-2} \{\; q^{1,1\;M}D^{0,2}\omega^{1,-1\;M} + 
q^{1,1\;M}D^{2,0}\omega^{-1,1\;M} +  \omega^{1,-1\;M}h^{1,3\;M} 
\nonumber \\
&&+ \omega^{-1,1\;M}h^{3,1\;M} + \omega^{-1,1\;M} \omega^{1,-1\;N}
\;h^{2,2\;[M,N]} + h^{2,2}\;\}\;, \label{haction}
\eea
where the involved potentials depend only on $q^{1,1\;M}$ and target 
harmonics and 
satisfy the target space constraints (\ref{1}) - (\ref{4}). To reveal the 
geometry hidden in these constraints we need their general 
solution, which is still unknown. At present we are only aware of 
some particular solution which will be presented below.  

The action (\ref{haction}) and constraints (\ref{1}) - (\ref{4}) 
enjoy a set of invariances. 

One of them is a mixture of reparametrizations in 
the target space (spanned by the involved superfields and target 
harmonics) and the transformations which are bi-harmonic 
analogs of hyper-K\"ahler 
transformations of Refs. \cite{{BGIO},{GIOSap}}. 

More interesting is another invariance which has no analog in the 
hyper-K\"ahler case and is a non-abelain and in general nonlinear 
generalization of the abelian gauge invariance (\ref{gauge}) 
\bea
\delta \omega^{1,-1\;M}  &=&
\left( D^{2,0}\delta^{MN} + 
\frac{\partial h^{3,1\;N}}{\partial q^{1,1\;M}}
\right) \sigma^{-1,-1\;N} - \omega^{1,-1\;L} \;
\frac{\partial h^{2,2\;[L,N]}}
{\partial q^{1,1\;M}}\;\sigma^{-1,-1\;N}\;, 
\nonumber \\
\delta \omega^{-1,1\;M}  &=&
- \left( D^{0,2}\delta^{MN} + \frac{\partial h^{1,3\;N}}{\partial 
q^{1,1\;M}} \right) \sigma^{-1,-1\;N} - \omega^{-1,1\;L} \;
\frac{\partial h^{2,2\;[L,N]}}{\partial q^{1,1\;M}}\;\sigma^{-1,-1\;N}\;, 
\nonumber \\
\delta q^{1,1\;M} &=& \sigma^{-1,-1\;N} h^{2,2\;[N,M]}\;. \label{gaugenab}
\eea
As expected, the action is invariant only with taking account of the 
integrability 
conditions (\ref{1}) - (\ref{4}). In general, these gauge 
transformations close with a field-dependent Lie bracket parameter: 
\be
\delta_{br} q^{1,1\;M} = \sigma^{-1,-1\;N}_{br} h^{2,2\;[N,M]}\;, \;\;
\sigma^{-1,-1\;N}_{br} = -\sigma^{-1,-1\;L}_1 \sigma^{-1,-1\;T}_2 
\frac{\partial h^{2,2\;[L,T]}}{\partial q^{1,1\;N}}\;.
\ee
We see that eq. (\ref{4}) guarantees the nonlinear closure of the 
algebra of gauge transformations (\ref{gaugenab}) and so it is a group 
condition similar to the Jacobi identity. 

Curiously enough, the 
gauge transformations (\ref{gaugenab}) augmented with the group 
condition (\ref{4}) 
are precise bi-harmonic counterparts of the two-dimensional version of 
basic relations of 
the so called Poisson nonlinear gauge theory 
which recently received some attention 
\cite{Ikeda}. The manifold $(q,u,v)$ can be 
interpreted as a kind of bi-harmonic extension of some Poisson 
manifold and the potential 
$h^{2,2\;[N,M]}(q,u,v)$ as a tensor field inducing the Poisson 
structure on this extension. 

It should be pointed out that it is the presence of the 
antisymmetric potential $h^{2,2\;[N,M]}$ that makes the considered case 
non-trivial and, in particular, the gauge invariance (\ref{gaugenab}) 
non-abelian. If 
$h^{2,2\;[N,M]}$ is vanishing, the 
invariance gets abelian and the constraints (\ref{2}) - (\ref{4}) 
are identically satisfied, while (\ref{1}) is solved by 
\be \label{hzero}
h^{1,3\;M} = \nabla^{0,2}\Sigma^{1,1\;M}(q,u,v),\;\;
h^{3,1\;M} = \nabla^{2,0}\Sigma^{1,1\;M}(q,u,v)\;,
\ee
with $\Sigma^{1,1\;M}$ being an unconstrained prepotential. Then, 
using the freedom with respect to the target space 
reparametrizations, one may entirely gauge away $h^{1,3\;M}, 
h^{3,1\;M}$, thereby reducing (\ref{haction}) to the dual 
action of twisted $(4,4)$ multiplets (\ref{dualq}). In the case of 
one triple $q^{1,1}, \omega^{1,-1}, \omega^{-1,1}$ the potential 
$h^{2,2\;[N,M]}$ vanishes identically, so the general action  
(\ref{genact}) for $n=1$ is actually equivalent to (\ref{dualq}). 
Thus {\it only for} $n\geq 2$ a new 
class of torsionful $(4,4)$ sigma models comes out. 
It is easy to see that the action (\ref{haction}) with non-zero 
$h^{2,2\;[N,M]}$ {\it does not} admit any duality transformation 
to the form with the superfields $q^{1,1\;M}$ only, because 
it is impossible to remove the dependence on $\omega^{1,-1\;N}, 
\omega^{-1,1\;N}$ from the 
equations for $q^{1,1\;M}$ by any  
local field redefinition with preserving harmonic analyticity. Moreover, 
in contradistinction to the constraints (\ref{qconstr}), these 
equations are compatible 
only with using the equation for $\omega$'s. So,  
the obtained system certainly does not admit in general any dual 
description in terms of twisted $(4,4)$ superfields. Hence, the 
left and right complex structures on the target space can be 
non-commuting. In refs. \cite{{Iv},{Iv2}} we have explicitly shown this 
non-commutativity for a particular class of the models in question. 
Now we briefly describe this example.

\vspace{0.2cm}
\noindent{\bf 4. Harmonic Yang-Mills sigma models.}
Here we present a particular solution to the constraints 
(\ref{1})-(\ref{4}). 
We believe that it shares many features of the general solution 
still to be found.

It is given by the following ansatz
\bea
h^{1,3\;N} &=& h^{3,1\;N} \;=\; 0\;; \; h^{2,2} \;=\; h^{2,2}(t,u,v)\;, \; 
\;t^{2,2} \;=\; q^{1,1\;M}q^{1,1\;M}\;; \nonumber \\
h^{2,2\;[N,M]} &=& b^{1,1} f^{NML} q^{1,1\;L}\;, \;b^{1,1} \;=\; 
b^{ia}u^1_iv^1_a\;, \; b^{ia} = \mbox{const}\;,
\label{solut}
\eea
where the real constants $f^{NML}$ are totally antisymmetric. 
The constraints (\ref{1}) - (\ref{3}) are identically satisfied 
with this ansatz, while (\ref{4}) is now none other than the 
Jacobi identity which tells us that the constants 
$f^{NML}$ are structure constants of some real semi-simple Lie 
algebra (the minimal possibility is $n=3$, the 
corresponding algebra being $so(3)$). Thus the $(4,4)$ sigma models 
associated with 
the above solution can be interpreted as a kind of Yang-Mills 
theories in the harmonic superspace. They provide the direct 
non-abelian generalization 
of the twisted multiplet sigma models with the action (\ref{dualq}) which 
are thus analogs of two-dimensional abelian gauge theory. 
The action (\ref{haction}) specialized to the  
case (\ref{solut}) is as follows
\bea
S^{YM}_{q,\omega} &=& 
\int \mu^{-2,-2} \{\; q^{1,1\;M} (\; D^{0,2}\omega^{1,-1\;M} + 
D^{2,0}\omega^{-1,1\;M} + b^{1,1} \;\omega^{-1,1\;L} \omega^{1,-1\;N}
f^{LNM}\; ) \nonumber \\
&&+ \;h^{2,2}(q,u,v) \} \;.\label{haction0} 
\eea
It is a clear analog of the Yang-Mills action in the first order formalism, 
$q^{1,1\;N}$ being an analog of the YM field strength and $b^{1,1}$ 
of the YM coupling constant. 

An interesting specific feature of this ``harmonic Yang-Mills theory'' is 
that the ``coupling constant'' $b^{1,1}$ is doubly charged (this is 
necessary for the balance of harmonic $U(1)$ charges). Since 
$b^{1,1} = b^{ia}u^1_iv^1_a$, we conclude that in the geometry of the 
considered class of $(4,4)$ sigma models a very essential role is played 
by the quartet constant $b^{ia}$. 
When $b^{ia} \rightarrow 0$, the non-abelian structure contracts into the 
abelian one and we reproduce the twisted multiplet action (\ref{dualq}). 
It also turns out that $b^{ia}$ measures the ``strength'' of 
non-commutativity of the left and right complex structures on the bosonic 
target. 

In the simplest case of $h^{2,2} = q^{1,1\;M} q^{1,1\;M}$ the physical 
bosons part of the action \p{haction0} (after fixing a WZ gauge with respect 
to \p{gaugenab} and a partial elimination 
of auxiliary fields) is given by the following expression  
\be   \label{bosact}
S_{bos} = \int d^2 x [dv] \left( {i\over 2}\;g^{0,-1\;iM}(x,v)\;
\partial_{--} q^{0,1\;M}_i (x,v) \right)\;.
\ee
Here the fields $g$ and $q$ are subjected to the harmonic differential
equations 
\bea  
&& \partial^{0,2} g^{0,-1\;iM} - 2 (b^{ka}v^{1}_a)\;f^{MNL} q^{0,1\;iN} 
g^{0,-1\;L}_k 
\;=\; 4i \partial_{++} q^{0,1\;iM} \nonumber \\
&& \partial^{0,2} q^{0,1\;iM} - 2f^{MLN} (b^{ka}v^1_a)
\;q^{0,1\;L}_k \; q^{0,1\;iN} 
\;=\; 0  \label{eq12} 
\eea
and $q^{0,1\;iM}$ contains the physical bosonic field as the first component 
in its $v$ decomposition 
$$
q^{0,1\;iM}(x,v) = q^{ia\;M}(x)v^1_a + ... \;.
$$ 
Solving eqs. \p{eq12}, one may express the involved functions in terms 
of $q^{ia\;M}(x)$, do the $v$ integration in \p{bosact} and find 
the explicit expressions for the bosonic metric and torsion potential. 
In \cite{{Iv},{Iv2}} this was done in the first non-vanishing order in 
$q^{ia\;M}$ and $b^{ia}$. This approximation proved to be sufficient 
for computing the relevant left and right complex structures (to the same 
order in fields), checking their properties and finding their mutual 
commutator. The latter was found to be non-vanishing and proportional to the 
coupling constant $b^{ia}$. 

Thus in the present case in the bosonic sector we encounter 
a more general geometry compared to the one associated with 
twisted $(4,4)$ multiplets. The basic characteristic 
feature of this geometry is the non-commutativity  of the left and right 
complex structures. 
It is easy to check this property also for general 
potentials $h^{2,2}(q,u,v)$ in (\ref{haction0}). It 
seems obvious that the general case (\ref{haction}), 
({\ref{1}) - (\ref{4}) reveals the same feature. 
Stress once more that this important property 
is related in a puzzling way to 
the non-abelian structure of the analytic superspace 
actions (\ref{haction0}), 
(\ref{haction}): the coupling constant $b^{1,1}$ 
(or the Poisson potential 
$h^{2,2\;[M,N]}$ in the general case) measures the strength of the 
non-commutativity of complex structures. 

\vspace{0.2cm}
\noindent{\bf 5. Outlook.}
The obvious problems for further study of the presented new class 
of $(4,4)$ sigma models are to compute the relevant 
metrics and torsions in a closed form and to try to utilize the 
corresponding manifolds as backgrounds for 
some superstrings. An interesting question is 
as to whether the constraints 
(\ref{1}) - (\ref{4}) admit solutions corresponding to the $(4,4)$ 
supersymmetric group 
manifold WZNW sigma models. The list of appropriate group manifolds 
has been given in \cite{belg}. The lowest dimension manifold 
with non-commuting left and right structures \cite{RSS} 
is that of $SU(3)$. Its dimension 8 coincides  
with the minimal bosonic manifold dimension at which a non-trivial 
$h^{2,2\;[M,N]}$ in (\ref{haction}) can appear. 

\vspace{0.2cm}
\noindent{\bf Acknowledgements.}
The author thanks the Organizers of Buckow Symposium for inviting him 
to participate and to give this talk. A partial support 
from the RFFR, grant 93-02-03821, and the INTAS, grants 93-127 and 94-2317, 
is acknowledged.


\begin{thebibliography}{99}
\bibitem{Zum} B. Zumino, Phys. Lett. {\bf B 87} (1979) 203
\bibitem{GIOShk} A. Galperin, E. Ivanov, V. Ogievetsky and E. Sokatchev, 
Commun. Math. Phys. {\bf 103} (1986) 515
\bibitem{GIO} A. Galperin, E. Ivanov and V. Ogievetsky, Nucl. Phys. 
{\bf B 282} (1987) 74
\bibitem{GIOSap} A.S. Galperin, E.A. Ivanov, V.I. Ogievetsky and 
E. Sokatchev, Ann. Phys. {\bf 185} (1988) 22
\bibitem{quat} A. Galperin, E. Ivanov and O. Ogievetsky, 
Ann. Phys.  
{\bf 230} (1994) 201
\bibitem{DVal} F. Delduc and G. Valent, Class. Quantum Grav. {\bf 10} 
(1993) 1201
\bibitem{GHR} S.J. Gates Jr., C. Hull and M. Ro\u{c}ek, Nucl. Phys. 
{\bf B 248} (1984) 157
\bibitem{HoPa} P.S. Howe and G. Papadopoulos, Nucl. Phys. 
{\bf B 289} (1987) 264
\bibitem{belg} 
Ph. Spindel, A. Sevrin, W. Troost and A. Van Proeyen, 
Phys. Lett. {\bf B 206} (1988) 71
\bibitem{Luest} E. Kiritsis, C. Kounnas and D. L\"ust, Int. J. Mod. Phys. 
{\bf A 9} (1994) 1361
\bibitem{RSS} M. Ro\u{c}ek, K. Schoutens and A. Sevrin, Phys. Lett. 
{\bf B 265} (1991) 303 
\bibitem{CHS} C. Callan, J. Harvey and A. Strominger, Nucl. Phys. {\bf B359} 
(1991) 611 
\bibitem{GI} O. Gorovoy and E. Ivanov, Nucl. Phys. {\bf B 381} 
(1992) 394  
\bibitem{IS} E. Ivanov and A. Sutulin, Nucl. Phys. {\bf B 432} (1994) 246
\bibitem{IK} E. A. Ivanov and S. O. Krivonos, J. Phys. A: Math. 
and Gen. {\bf 17} (1984) L671 
\bibitem{BRL} T. Buscher, U. Lindstr\"om and M. Ro\u{c}ek, Phys. Lett. 
{\bf B 202} (1988) 94
\bibitem{DS} F. Delduc and E. Sokatchev, Int. J. Mod. Phys. {\bf B 8} 
(1994) 3725
\bibitem{RIL} U. Lindstr\"om, I.T. Ivanov and M. Ro\u{c}ek, Phys. Lett. 
{\bf B 328} (1994) 49
\bibitem{Iv} E. Ivanov, Preprint ESI-196 (1995), 
JINR-E2-95-53, hep-th/9502073 (Phys. Rev. D, in press)
\bibitem{Iv2} E. Ivanov, Phys. Lett. {\bf B 356} (1995) 239
\bibitem{GK} S. James Gates, Jr. and S.V. Ketov, Preprint ITP-UH-15/95, 
UMDEPP 95-116, hep-th/9504077 
\bibitem{Ikeda} N. Ikeda, Ann. Phys. {\bf 235} (1994) 435; 
P. Schaller and T. Strobl, Mod. Phys. Lett. 
{\bf A 9} (1994) 3129 
\bibitem{BGIO} J.A. Bagger, A.S. Galperin, E.A. Ivanov and V.I. Ogievetsky, 
Nucl. Phys {\bf B 303} (1988) 522
 


\end{thebibliography}
\end{document}